\documentclass[%
 reprint,
 amsmath,amssymb,
 aps,
 prl
]{revtex4-2}

\usepackage{graphicx}
\usepackage{dcolumn}
\usepackage{bm}

\usepackage{comment}
\usepackage{balance}
\begin{document}

\preprint{APS/123-QED}

\title{Ultracold Neutron Sources: An Essentially Enhanced Optimal Operation Temperature for Ortho-Deuterium-Based Superthermal Converters }

\author{Erik Walz}
\affiliation{Heinz Maier-Leibnitz Zentrum (MLZ), Technical University Munich, 85748 Garching, Germany}
\email{Corresponding author: erik.walz@frm2.tum.de}

\author{Christoph Morkel}
\affiliation{Technical University Munich, TUM School of Natural Sciences, Physics Department, 85748 Garching, Germany}

\date{\today}

\begin{abstract}
There is a longstanding dogma in the physics of ultracold neutrons (UCN): The optimal temperature of ortho-deuterium based UCN-sources must not exceed $5\,\mathrm{K}$. This value has been established by the pioneering work of Golub et al. decades ago, utilizing previously established cross sections for solid ortho-deuterium. But recently, an improved calculation of the crucial 1-phonon up-scattering cross section - using the so-called corrected Incoherent Approximation - leads to a new upper limit for the optimal operation temperature  of ortho-deuterium based UCN sources. After a discussion of the relevant cross sections of the \(\mathrm{D}_2\)-molecule we conclude that the optimal temperature lies between $10\,\mathrm{K}$ and $12\,\mathrm{K}$, a factor 2 higher than estimated earlier. This offers significantly relaxed cryogenic conditions for UCN experiments with solid ortho-deuterium and especially for the operation of ortho-deuterium based ultracold neutron sources. 
\end{abstract}

\maketitle


\section{\label{sec:Intro} Introduction} 
Research reactors around the world are equipped with a source of ultracold neutrons (UCN) using solid ortho-deuterium (o\text{-}\ensuremath{\mathrm{D}_2}) as an efficient superthermal converter. These UCN-sources are operated at low temperature near \( T \approx 5\,\mathrm{K} \), as deduced decades ago by \citet{Yu1986}. These authors applied the widely used Incoherent Approximation (I.A.) for 1-phonon up-scattering \cite{Gurevich1968}, treating the prevailing coherent scattering in o-\(\mathrm{D}_2\) also as incoherent scattering. Our improved calculation of  \( T^* \) - the optimal operation temperature - uses the so-called corrected Incoherent Approximation for 1-phonon up-scattering \cite{Placzek1955, Doege2021}. This approximation takes the interference effects due to coherent scattering of the o-\(\mathrm{D}_2\) molecule into account. This scattering is most important in an o-\(\mathrm{D}_2\) crystal with a ratio larger than 4 between coherent and incoherent cross section.

\section{\label{sec:Cross Sections} The relevant cross sections}

The calculation of the optimal operation temperature \( T^* \) requires a careful discussion of the relevant cross sections for UCN-production and -losses in a solid o\text{-}\ensuremath{\mathrm{D}_2} moderator.

\subsection{\label{sec:production cross sections}UCN-production in o\text{-}\ensuremath{\mathrm{D}_2}}

As shown by \citet{Frei2010} the UCN-production cross section \( \sigma^{+1\text{ph}}_{E_0} \) (down scattering of cold neutrons by phonon excitation) is given at low temperature as:

\begin{equation}
\sigma^{+1\text{ph}}_{E_0} = 2 \, \sigma_{\mathrm{SC}} \, \frac{m_n}{m_{\mathrm{D}_2}} \, 
\frac{E_0^{3/2} \, E_{\mathrm{UCN}}^{3/2}}{(k \, \Theta_D)^3} \,
\langle n + 1 \rangle \, e^{-2W(k_0)} \;,
\label{eq:prod_cross_section}
\end{equation}

while a more involved calculation is given by Golub \cite{Golub1983, Yu1986}.

In Eq.~(\ref{eq:prod_cross_section}),  \( \sigma_{\mathrm{SC}} \) is the total scattering cross section (\( \sigma_{\mathrm{SC}} = 27.5\,\mathrm{barn} \) for o\text{-}\ensuremath{\mathrm{D}_2}), \(\frac{m_n}{m_{\mathrm{D}_2}} = \frac{1}{4}\) is the mass ratio between neutron and \ensuremath{\mathrm{D}_2} molecule, \( E_0 \) is the initial energy of the cold neutron (\( E_0 \approx 2\,\mathrm{meV} \)), and 
\( E_{\mathrm{UCN}} \) is the upper energy limit of the UCN yield of the source (\( E_{\mathrm{UCN}} \approx 150\,\mathrm{neV} \)). It should be mentioned that for cold neutrons with energies of a few \(\mathrm{meV}\), the Incoherent Approximation (used in Eq.~(\ref{eq:prod_cross_section})) applies well \cite{Placzek1951} with \( \Theta_D \) the Debye temperature of the moderator (\( \Theta_D = 110\,\mathrm{K} \) for  o\text{-}\ensuremath{\mathrm{D}_2} \cite{Souers1986}). The Bose factor \( \langle n + 1 \rangle \) for phonon excitation is nearly temperature independent, with \( \langle n + 1 \rangle = \left( e^{\frac{E_0}{kT}} - 1 \right)^{-1} + 1 \approx 1 \) at low temperatures. Using \( 2W_{k_0} = \tfrac{\langle u^2 \rangle}{3} k_0^2 \), with
\( \langle u^2 \rangle = 0.25\,\text{\AA}^2 \) for solid \( \mathrm{D}_2 \) at
\( T \to 0 \) and \( k_0(E_0) \approx 1\,\text{\AA}^{-1} \),
the Debye--Waller factor is found to be nearly unity. Hence, the cross section for UCN-production is nearly temperature independent. This is different for UCN-losses \( \sigma^{-1\text{ph}}_{E_{\mathrm{UCN}}} \) in o-\(\mathrm{D}_2\), which is the main issue of this paper.

\subsection{\label{sec:loss cross sections}UCN-losses in o\text{-}\ensuremath{\mathrm{D}_2} (Incoherent Approximation)}

Opposed to UCN down-scattering, there is a loss channel due to UCN up-scattering by 1-phonon annihilation processes in o\text{-}\ensuremath{\mathrm{D}_2}. The corresponding cross-section is given in a first approach by the widely used Incoherent Approximation (I.A.) \cite{Gurevich1968}, which can be expressed in the limit \( T \to 0 \) and \( E_{\mathrm{UCN}} \to 0 \) as \cite{Stepanov1975}:

\begin{equation}
\sigma^{-1\text{ph}}_{E_{\mathrm{UCN}}} = \sigma_{\mathrm{SC}} \, \frac{45}{8} \, \sqrt{\pi} \, \zeta\left( \frac{7}{2} \right) \,
\frac{m_n}{m_{\mathrm{D}_2}} \left( \frac{kT}{E_{\mathrm{UCN}}} \right)^{1/2} \left( \frac{T}{\Theta_D} \right)^3 \;,
\label{eq:ucn_loss_cross_section}
\end{equation}

with the Riemann zeta function \( \zeta\left( \frac{7}{2} \right) = 1{.}1267 \) \cite{Abramowitz1966}. This simple expression already displays the two important features of the UCN 1-phonon up-scattering cross section. Firstly, there is a characteristic \( 1/v_{\mathrm{UCN}} \) behavior, which makes UCN up-scattering to the leading loss channel at low neutron energies. Secondly, there is a pronounced \( T^{7/2} \) dependence of the cross-section at \( T \to 0 \). To avoid a large amount of UCN up-scattering it is therefore necessary to freeze out the phonon population in the moderator and to cool it down to temperatures \( T^* \) as low as reasonably achievable. This temperature has first been estimated by \citet{Yu1986} and \citet{Golub1991} to be \( T^* \approx 4\,\mathrm{K} \) for o\text{-}\ensuremath{\mathrm{D}_2} (indicated in Fig.~\ref{fig:golub}). An improved expression for the Stepanov Eq.~(\ref{eq:ucn_loss_cross_section}) has been given by \citet{Gurevich1968}, which was used by \citet{Doege2021} to calculate numerical values for \( \sigma^{-1\text{ph}}_{E_{\mathrm{UCN}}} \) using the I.A. (see Tab. \ref{tab:sigmas}).

\subsection{\label{sec:corrected IA}UCN losses: Corrected Incoherent Approximation}

The application of the I.A. in o-\(\mathrm{D}_2\) is problematic, although widely used, because of interference effects due to prevailing coherent phonon up-scattering in the moderator. This is best seen in the formula given by \citet{Placzek1955} in 1955:

\begin{equation}
\sigma^{-1\text{ph}} = \sigma_{\mathrm{SC}} \, S^{\mathrm{inc}} \left( 1 + \frac{\sigma^{\mathrm{coh}}}{\sigma_{\mathrm{SC}}} \cdot \frac{\delta S}{S^{\mathrm{inc}}} \right),
\label{eq:pvH}
\end{equation}
with \( \delta S = S^{\mathrm{coh}} - S^{\mathrm{inc}} \), where \( S^{\mathrm{coh}} \) and \( S^{\mathrm{inc}} \) are the coherent and incoherent integrated scattering law for  1-phonon annihilation (phonon up-scattering of UCN), respectively. The weighted correction factor \( \frac{\delta S}{S^{\mathrm{inc}}} \) (\( \frac{\sigma^{\mathrm{coh}}}{\sigma_{\mathrm{SC}}} = 0.815\) for o\text{-}\ensuremath{\mathrm{D}_2}) takes differences between coherent and incoherent scattering into account and has been calculated according to \citet{Placzek1955} for o\text{-}\ensuremath{\mathrm{D}_2} by \citet{Doege2021}. The results are given in Tab. \ref{tab:sigmas} for different temperatures together with new values for para\text{-}\ensuremath{\mathrm{H}_2}, which is a purely coherent scatterer (\( \sigma^{\mathrm{coh}} = 7.03\,\mathrm{barn/molecule} \)).

\begin{table}[h]
\caption{\label{tab:sigmas}Phonon up-scattering cross sections for solid o\text{-}\ensuremath{\mathrm{D}_2} (melting temperature $T_m = 18.7\,\mathrm{K}$, left part) and p\text{-}\ensuremath{\mathrm{H}_2} ($T_m = 13.8\,\mathrm{K}$, right part). The subscript I.A. marks the Incoherent Approximation \citep{Gurevich1968,Doege2021}, whereas PvH denotes the results from Eq.~(\ref{eq:pvH}) with $\frac{\delta S}{S^{\mathrm{inc}}}$ given by \citep{Placzek1955} and calculated by \citep{Doege2021,Placzek1955}. It is evident that the corrections to the I.A. results are as large as factors 3 to 5 and even larger for p\text{-}\ensuremath{\mathrm{H}_2} ($\sigma_{\mathrm{coh}} / \sigma_{\mathrm{sc}} = 1.0$). All values are given per molecule, scale with $1/v_{\mathrm{ucn}}$ and refer to $v_{\mathrm{ucn}} = 10\,\mathrm{m/s}$.}
\begin{ruledtabular}
\begin{tabular}{c|cc||cc}
\multicolumn{1}{c}{} 
& \multicolumn{2}{c||}{o\text{-}\ensuremath{\mathrm{D}_2}} 
& \multicolumn{2}{c}{p\text{-}\ensuremath{\mathrm{H}_2}} \\
$T$ (K) 
& $\sigma^{-1\text{ph}}_{\text{I.A.}}$ 
& $\sigma^{-1\text{ph}}_{\text{PvH}}$ 
& $\sigma^{-1\text{ph}}_{\text{I.A.}}$ 
& $\sigma^{-1\text{ph}}_{\text{PvH}}$ \\
\hline
5  & 0.19  & 0.04  & 0.12  & 0.006 \\
10 & 1.99  & 0.63  & 1.19  & 0.18  \\
13 & 4.68  & 1.83  & 2.72  & 0.76  \\
15 & 7.39  & 3.40  &       &       \\
18 & 12.9  & 6.70  &       &       \\
\end{tabular}
\end{ruledtabular}
\end{table}

\subsection{\label{sec:loss by absorption}UCN losses: Absorption}

Another loss channel for UCN is absorption from \ensuremath{\mathrm{D}_2} and some unavoidable parasitic absorbers as para\text{-}\ensuremath{\mathrm{D}_2} (p\text{-}\ensuremath{\mathrm{D}_2}) and hydrogen (\ensuremath{\mathrm{H}_2}). As an approximation, we have: 

\begin{equation}
\sigma^{\mathrm{abs}}_{\mathrm{D}_2, \mathrm{tech}}(v) = c_0 \, \sigma^{\mathrm{abs}}_{\mathrm{D}_2}(v) + (1 - c_0) \, \sigma^{(p \to 0)}_{p -\mathrm{D}_2} + c_{\mathrm{H}_2} \, \sigma^{\mathrm{abs}}_{\mathrm{H}_2}(v),
\label{eq:absorption_tech_D2}
\end{equation}

where \( c_0 = 0.975 \) is the number concentration for technical o\text{-}\ensuremath{\mathrm{D}_2} \cite{Doege2019}, \( \sigma^{\mathrm{abs}}_{\mathrm{D}_2} = 0.23\,\mathrm{barn/molecule} \) and \( \sigma^{(p \to 0)}_{p - \mathrm{D}_2} = 27.1\,\mathrm{barn/molecule} \) \cite{Liu2000} from UCN up-scattering due to the rotational relaxation of  p\text{-}\ensuremath{\mathrm{D}_2} (rotational quantum number \( J' = 1 \)) to o\text{-}\ensuremath{\mathrm{D}_2} (\( J = 0 \), ground state) with \( \Delta E_{J'J} = +7.4\,\mathrm{meV} \) transferred to the UCN. Then there is parasitic absorption from \ensuremath{\mathrm{H}_2} of \( \sigma_{\mathrm{H}_2} = c_{\mathrm{H}_2} \cdot 146\,\mathrm{barn/molecule} \) with \( c_{\mathrm{H}_2} = 1 \times 10^{-3} \) \cite{Sears1984}. All these cross sections scale with \( 1/v_{\mathrm{UCN}} \) and are given here for \( v_{\mathrm{UCN}} = 10\,\mathrm{m/s} \). Therefore, according to Eq.~(\ref{eq:absorption_tech_D2}) the total absorption in technical o\text{-}\ensuremath{\mathrm{D}_2} results in: 
\( \sigma^{\mathrm{abs}}_{\mathrm{D}_2, \mathrm{tech}}(v = 10\,\mathrm{m/s}) = 0.22 + 0.68 + 0.15 = 1.05\,\mathrm{barn/molecule} \)
as level of absorption, which is independent of temperature \cite{Liu2000} and unavoidable in solid o\text{-}\ensuremath{\mathrm{D}_2} based UCN-sources. 

\section{\label{sec:operation temperature}The optimal operation temperature}

\begin{figure}[ht]
\centering
\includegraphics[width=1.0\columnwidth]{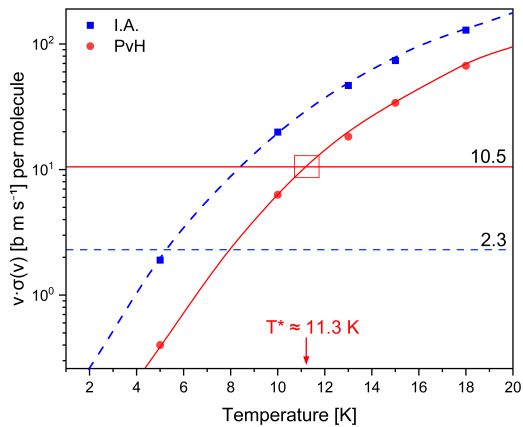}
\caption{\label{fig:golub}
Comparison of $T$-independent absorption (horizontal lines: $\sigma^{\mathrm{abs}}_v \cdot v = \text{const.}$) and $T$-dependent phonon up-scattering $\sigma^{-1\text{ph}}_v \cdot v$ (guide to the eye) for solid o\text{-}\ensuremath{\mathrm{D}_2}. The blue lines reproduce the I.A. results of \citet{Golub1983}; the red lines display the new calculation performed here, showing an optimal temperature \( T^* \approx 11\,\mathrm{K} \) for the corrected I.A. and the absorption cross section of technical o\text{-}\ensuremath{\mathrm{D}_2}.}
\end{figure}

\citet{Golub1983} have first calculated the operation temperature of solid \ensuremath{\mathrm{D}_2} based UCN-sources by comparing T-dependent phonon up-scattering with T-independent absorption. However, the phonon up-scattering calculation following the I.A. is known to yield questionable results \cite{Doege2021,Liu2010}. Applying interference corrections to the I.A. (see Eq.~(\ref{eq:pvH}) \cite{Doege2021,Doege2019}) leads to a much lower up-scattering cross section in solid o\text{-}\ensuremath{\mathrm{D}_2} at low temperatures, displayed here in Fig.~\ref{fig:golub}, adapted from \citet{Golub1991}. As a consequence, the temperature \( T^* \), at which phonon up-scattering freezes out below the absorption limit is no longer \( 4\,\mathrm{K} \) \cite{Golub1983}, but reaches \( T^* \approx 11\,\mathrm{K} \) as optimal operation temperature of these UCN-sources with respect of the absorption cross sections of technical o-\(\mathrm{D}_2\).

Earlier I.A. results (dashed lines in Fig.~\ref{fig:golub} \cite{Golub1991}) are about a factor of 2 too high. Freezing out phonons below \( T^* \approx 11\,\mathrm{K} \) does not reduce UCN-losses further, because T-independent absorption processes dominate below \( T^* \). Parasitic absorption by p\text{-}\ensuremath{\mathrm{D}_2} and \ensuremath{\mathrm{H}_2}  has been included here according to Eq.~(\ref{eq:absorption_tech_D2}). Our new calculation of \( T^* \) exceeds an earlier Monte-Carlo simulation of \citet{Liu2010} by about \( 4\,\mathrm{K} \) and is shown here to be independent of the UCN-velocity. It is corroborated by a conjecture of \citet{Lavelle2010} and \citet{Doege2019}. They found experimental evidence that a variety of \ensuremath{\mathrm{D}_2} based UCN-sources in fact do show a marked decrease of the UCN yield only in the region beyond \( 12\,\mathrm{K} \). No such optimal operation temperature can be deduced in the case of solid \(\mathrm{H}_2 \), due to the extraordinarily high absorption of hydrogen. A tentative increase of the p\text{-}\ensuremath{\mathrm{D}_2} concentration to 3\% raises the absorption level by 0.14~barn/molecule. Such an additional absorption would shift $T^\ast$ to a slightly higher value of about 11.5~K. In general, however, the p\text{-}\ensuremath{\mathrm{D}_2} concentration should be kept as low as possible in practice.

\section{\label{sec:conclusion}Conclusion}

An improved calculation of the two UCN loss channels in o\text{-}\ensuremath{\mathrm{D}_2} (T-independent absorption versus T-dependent phonon up-scattering) using the corrected Incoherent Approximation for combined coherent and incoherent 1-phonon up-scattering enabled us to derive a substantially enhanced optimal operation temperature \( T^* \approx 11\,\mathrm{K} \) for solid o\text{-}\ensuremath{\mathrm{D}_2} cryo-moderators. This is about a factor 2 higher than estimated earlier and offers relaxed cryogenic conditions for UCN experiments with solid o\text{-}\ensuremath{\mathrm{D}_2} and - more important - for the operation of these UCN-sources. Moreover, we conclude that frequently discussed temperature gradients within a \(\mathrm{D}_2\)-source can be tolerated as long as the temperature itself does not exceed \( T^* \).

\section{\label{sec:Ackn}Acknowledgments}
The authors thank Andreas Frei and Robert Golub for fruitful discussions.

\bibliography{UCN_libary}
\end{document}